\documentclass[11pt]{llncs}        
 
\usepackage{fullpage}    
\usepackage{proof}
\usepackage{times}
\usepackage{rotating}
\usepackage{amssymb}
\usepackage{amstext}
\usepackage{pst-node,pst-tree}

\newpsobject{showgrid}{psgrid}{subgriddiv=1,griddots=0,gridlabels=6pt}

\newcommand{\myparagraph}[1]{\vspace*{2mm}\noindent \textbf{#1}}
                                  
\newcommand{\Vampire}{Vampire}

\newcommand{\Gandalf}{Gandalf}

\newcommand{\Prolog}{Prolog}

\newcommand{\inc}{~~\= \+ \kill}
\newcommand{\dinc}{\inc\inc}

\newcommand{\emptyclause}{\Box}

\begin{document}

\mainmatter

\title{Resolution-based Query Answering for Semantic Access to Relational Databases: A Research Note}

\author{Alexandre Riazanov}

\institute{
    \email{alexandre.riazanov@gmail.com}\\ \vspace{1cm}
    March 29, 2009
}

\maketitle

\abstract{
  We address the problem of semantic querying of relational 
  databases (RDB) modulo knowledge
  bases using very expressive knowledge representation
  formalisms, such as full first-order logic or its
  various fragments. We propose to use a first-order logic (FOL)
  reasoner for computing
  \emph{schematic answers} to deductive queries, with the \emph{subsequent
  instantiation} of these schematic answers using a conventional
  relational DBMS. In this research note, we outline the main
  idea of this technique -- using \emph{abstractions of databases}
  and \emph{constrained clauses} for deriving schematic answers.
  The proposed method can be directly used with \emph{regular RDB}, including \emph{legacy
  databases}. Moreover, we propose it as a potential basis for 
  an efficient Web-scale semantic search technology.
}


                 \section{Introduction.}


\subsection{Settings and motivation.} 

Consider the following
scenario. Suppose we have one\footnote{In principle, our approach
can be extended to multiple heterogeneous and distributed databases, 
but in this research note we assume, for simplicity, that we are dealing with 
just one DB.
} relational database (RDB),
one or more expressive knowledge bases (KB) for domains 
to which the data
in the RDB is related (e.~g., rule bases and/or ontologies), 
and, optionally, some mapping between 
the RDB schema and the logical language of the domains,
i.~e., a logical description of the relations in the RDB to link 
them to the concepts and relations defined by the 
knowledge bases. We would like to be able to formulate queries
logically and answer them w.~r.~t. the knowledge bases and the RDB treated 
virtually as a collection of ground atomic facts 
(e.~g., by viewing each table row as a separate ground fact).
\emph{To make this process efficient, we would like to use 
the modern RDB technology as much as possible by delegating 
as much work as possible to the RDBMS hosting the database}.

We propose a method to implement this scenario,
based on the use of \emph{resolution for incremental transformation of
semantic queries into sequences of SQL queries} that can be directly
evaluated on the RDB, and whose results provide answers to
the original queries.

We envisage two main applications for the proposed 
technology. 

\myparagraph{Enhancing the interface to conventional relational databases.}
Flexible querying of conventional RDBs by non-programmer users is very 
problematic because real-life enterprise databases often 
have complex designs. Writing a correct query requires 
good understanding of technical details of the DB schema,
such as table and attribute names, foreign key relationships, 
nullable fields, etc. 
So most of RDB querying by non-programmer users
is done with preprogrammed parameterised queries, usually represented 
as forms of various kinds.

Even when special methodologies are used, like Query-by-Example 
(see, e.~g. \cite{Ramakrishnan:Database_Management_Systems:2003}), 
that allow to hide some of the complexities of SQL and database designs 
from the end users, one important inherent limitation remains 
in force. Whereas mapping some domain concepts to the RDB schema elements
may be easy, many other concepts may be much more difficult to map.
For example, it is easy to select instances of the concept 
``student'' if there is a table explicitly storing all students,
but if the user wants to extract a list of all members
of a department in a university, he may have to \emph{separately} query 
different tables storing information about students, faculty and 
support staff (assuming that there is no table specifically storing
members of all these kinds), and then create a union of the results.

This example exposes well the root of the problem: mapping some 
domain concepts to the data is difficult because it requires 
\emph{application of the domain knowledge}. 
In the example, the involved piece of
domain knowledge is the fact that students, faculty and support 
staff are all department members, and the user has to apply it 
manually to obtain the required results. 

\emph{Semantic querying} is based on automatic application of domain 
knowledge formalised in the form of, e.~g., rules and 
ontological axioms.
In this approach, DB programmers ``semantically document'' their DB designs 
by providing an explicit correspondence between the RDB schemas and
domain terminologies, e.~g., in the form of logical axioms.
This alone allows an end user to formulate queries directly
in the terminology of the domain, without even a slightest
idea about how the underlying RDBs are structured\footnote{This does not
alleviate the need for convenient query interfaces, but they are outside 
the scope of this research note.}. 
However, the biggest advantage comes from the fact that 
reasoning w.~r.~t. additional, completely external KBs 
can be employed to generate and
justify some answers, which makes querying not just \emph{semantic},
as in \cite{Rishe:SemSQL}, but also \emph{deductive}. 
In our current example, the user can provide,
as a part of the query, some KB that links the relations of being a  
department member, being a student in the department, etc. 
In some application contexts, it is important to be able to use rather 
expressive KBs for such purposes. Rule-based KBs and expressive
DL ontologies are of a special interest, especially in combination.

Database application development can also benefit from 
the proposed technology. Currently, it is often difficult 
to change the structure of an existing corporate database if there are 
many applications relying on the particular schema being used. 
Our scenario suggests an attractive alternative -- applications
that only query the DB semantically need not be rewritten
when the DB design changes, provided that a suitable semantic
mapping can be written for the new design.
In particular, new DBs can be added without changing the applications
that will query them. With such decoupling of applications from the DB
designs, it may also be easier to outsource the development
of applications because exposing the details of DB designs to application
developers becomes unnecessary.

\myparagraph{Web-scale semantic search.} The Semantic Web is expected
to contain a lot of data in the form of RDF and OWL 
descriptions referring to various formalised vocabularies -- ontologies. 
In some cases the expressivity of RDF(S) and OWL
may not be enough and knowledge bases in other formalisms,
e.~g., RuleML \cite{Boley+:RuleMLDesignRationale:SWWS:2001,Boley+:RuleML:URL}, 
RIF \cite{Hawke:RIFCharter:URL}
or SWRL \cite{W3C:SWRL:URL}, have to be used to capture more complex
dependencies between domain concepts and relations, thus
making the data descriptions sufficiently semantically rich.

The utility of the Semantic Web data will strongly depend on 
how easily and how efficiently users and agents can query it.
Roughly speaking, we need to \emph{query extremely large volumes of 
highly distributed data modulo expressive knowledge bases}, so that not only 
direct answers based on the stored data are returned, 
but also implied answers that can only be obtained
by reasoning.

The approach proposed here may be a part of a solution to this problem:
large sets of RDF triples and OWL data descriptions (coming from 
Semantic Web documents)
can be loaded into a relational database and then queried 
deductively modulo the relevant knowledge bases. Different DB layouts
can be used, depending on the nature of the data being loaded.
For example, if we load an OWL ABox, we can have a separate one-column 
table for keeping instances of each class and, similarly, a separate 
two-column table for keeping assertions of each property\footnote{
These is the scheme used in all examples throughout the research note.
}. 
Loading data descriptions into an RDB is a linear
operation, so it is unlikely to become a real performance bottleneck.
Moreover, we can start producing answers even before the data is fully 
loaded. So the efficiency of such a scheme depends mostly on how
efficiently the deductive querying on the RDB can be done.

Just like text-based Web search engines do not indiscriminately 
scan all the accessible documents each time a new query is processed, 
semantic search systems cannot examine all accessible 
data descriptions in every retrieval attempt. 
Instead, some form of indexing is necessary
that would allow to avoid downloading data that is irrelevant 
to a specific query, and would focus the processing on the sets of assertions
that are likely to contribute to some answers to the query. 
We will show that the core feature of our approach to deductive 
querying of RDB -- incremental query rewriting -- suggests a natural way of 
semantically indexing distributed data sources.

\subsection{Outline of the proposed method.} To implement 
the target scenario,
we propose to use a first-order logic reasoner in combination
with a conventional RDBMS, so that the reasoner does the ``smart''
part of the job, and the RDBMS is used for what it is best at 
-- relatively simple processing of large volumes of 
relational data by computing table joins.
Roughly, the reasoner works as a query
preprocessor. It accepts a semantic query, the relevant knowledge bases 
and a semantic mapping for a DB as its input, and
generates a (possibly infinite) number of expressions which we
call \emph{schematic answers}\footnote{In earlier versions of this research note we used
the term \emph{generic answers}, which clashes with the classification proposed 
in \cite{BurhansShapiro:JAL:AnswerClasses:2007}.
}, that can be easily converted into SQL
queries.
These SQL queries are then evaluated on the DB with the help of 
the RDBMS. The union of the results for these SQL queries 
contains all answers to the original deductive query.

\begin{figure*}
\begin{center}
\psset{xunit=.07cm,yunit=.05cm,dash=3pt 4pt}
\begin{pspicture}(-10,25)(150,105)

\pscircle(0,80){5pt}
\pnode(0,75){n1}
\pnode(5,65){n2}
\pnode(-5,65){n3}
\ncline{n1}{n2}
\ncline{n2}{n3}
\ncline{n3}{n1}
\rput(0,60){User/client}
\rput(0,55){code}

\psframe(25,35)(45,90)
\rput(35,65){Reasoner}

\pnode(45,80){n1}
\pnode(70,80){n2}
\ncline{->}{n1}{n2}
\rput(57,89){general}
\rput(57,84){solutions}

\pnode(15,80){n1}
\pnode(25,80){n2}
\pnode(7,72){n3}
\ncline{->}{n1}{n2}
\ncline{n3}{n1}
\rput(18,84){Query}

\pnode(15,60){n1}
\pnode(25,60){n2}
\ncline{->}{n1}{n2}
\rput(19,64){KBs}

\pnode(15,40){n1}
\pnode(25,40){n2}
\pnode(7,50){n3}
\ncline{->}{n1}{n2}
\ncline{n3}{n1}
\rput(16,35){RDB}
\rput(16,30){abstraction}

\psframe(70,55)(90,90)
\rput(80,75){SQL} 
\rput(80,68){generator}

\pnode(45,60){n1}
\pnode(70,60){n2}
\ncline[linestyle=dashed]{->}{n1}{n2}
\rput(57,64){constraints}

\pnode(90,80){n1}
\pnode(120,80){n2}
\ncline{->}{n1}{n2}
\rput(105,84){SQL queries}

\pnode(90,60){n1}
\pnode(120,60){n2}
\ncline[linestyle=dashed]{->}{n1}{n2}
\rput(105,69){SQL queries}
\rput(105,64){for constraints}

\pnode(120,45){n1}
\pnode(45,45){n2}
\ncline[linestyle=dashed]{->}{n1}{n2}
\rput(83,48){feedback on constraint satisfiability}

\psframe(120,35)(140,90)
\rput(130,65){RDBMS}

\pnode(130,90){n1}
\pnode(130,100){n2}
\pnode(15,100){n3}
\pnode(7,92){n4}
\ncline{n2}{n1}
\ncline{n2}{n3}
\ncline{->}{n3}{n4}
\rput(80,104){Query answers}

\psframe(120,25)(140,35)
\rput(130,30){RDB}

\end{pspicture}
                   \caption{Architecture for deductive query answering
                         \label{fig:architecture}}

\end{center}
\end{figure*}
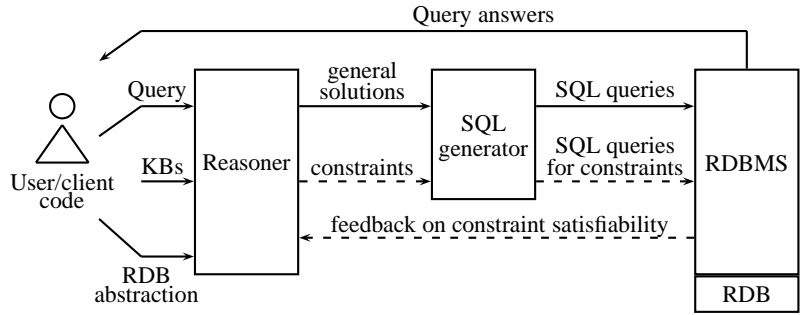

This idea can be implemented with a relatively simple
architecture as shown in Figure~\ref{fig:architecture}.
The architecture introduces two main modules -- a reasoner for finding schematic
solutions and an SQL generator to turn these solutions into SQL queries.
We also assume that some off-the-shelf RDBMS is used to answer the SQL queries.
All three components (can) work in parallel: while the reasoner 
searches for another schematic answer, the SQL generator can process some previous
general solutions and the RDBMS can generate instances for some earlier 
general solutions and communicate them to the user.

Optionally, the reasoner may try to prune the search space by checking
certain constraints over the RDB (details will be provided in 
Section~\ref{sec:constraints}). These constraints are also converted
into SQL queries and sent to the RDBMS for evaluation. 
The results of the evaluation ($'satisfiable'$ or $'unsatisfiable'$) 
are sent back to the reasoner which can use the absence of solutions 
for a constraint as a justification for suppressing certain inferences.


The rest of this research note is structured as follows.
In Section~\ref{sec:main_idea} we introduce the method
intuitively. 
In Section~\ref{sec:soundness_and_completeness}
we provide a minimal mathematical justification of usability
of our approach. More specifically, we demonstrate soundness 
and completeness of some standard resolution-based calculi
for rewriting semantic queries into sequences of schematic answers.
In Section~\ref{sec:constraints} we describe one optimisation specific 
to schematic answer search. 
In Section~\ref{sec:indexing_with_abstractions} we briefly discuss
how semantic indexing can be done using data abstractions,
in the context of Web-scale retrieval.
In Section~\ref{sec:sql-generation}
we provide an algorithm for converting the logical representation of
schematic answers into SQL. 
Finally, Sections~\ref{sec:related_work} and~\ref{sec:future_work}
briefly describe some related and future work.


                 \section{Informal method description.}
                    \label{sec:main_idea}

We \emph{model} an RDB as a finite set of ground atomic formulas, so that RDB 
table names provide the predicates, and rows are conceptually treated
as applications of the predicates to the row elements. In the example below,
we have a table $takesCourse$ from a University DB, 
keeping information about which student takes which course, 
whose rows are mapped to a set of facts.
\begin{center}
\begin{tabular}{l|c|c|cl} 
\large ~~takesCourse~~  & \large {\bf ~~student~~} & \large {\bf ~~course~~} \\ 
\normalsize
                & s1      & c1      & ~~~~~~~~~~~~~~~~~~~~$\longrightarrow$~~~~~~~~~~~~~~~~~~~~ & takesCourse(s1,c1)\\
                & s2      & c2      & ~~~~~~~~~~~~~~~~~~~~$\longrightarrow$~~~~~~~~~~~~~~~~~~~~ & takesCourse(s2,c2)\\
                & s3      & c3      & ~~~~~~~~~~~~~~~~~~~~$\longrightarrow$~~~~~~~~~~~~~~~~~~~~ & takesCourse(s3,c3)\\
                & $\ldots$ & $\ldots$ & $\ldots$
\end{tabular}
\end{center}

Before we proceed with more important things, note that in all our examples in this research note,
the data is assumed to be a relational representation of some DL ABoxes. This is done
not to clutter the presentation of the main ideas with RDB schema-related details. 
In particular, there is no need for a special RDB-to-KB mapping because 
the RDB tables directly correspond to concepts and properties.It bears repeating that
this assumption is made \emph{only to simplify the presentation} -- our approach is
applicable to any RDBs, including legacy ones, as long as their design allows
reasonable semantic mapping.

\vspace*{2mm}
Now, suppose we are trying to answer a query over our RDB deductively,
e.~g., modulo some KB. 

\myparagraph{Naive approach as a starting point.} Hypothetically, we can explicitly
\emph{represent the DB as a collection of ground atomic facts}
and use some resolution-based FOL reasoner supporting query answering, 
e.g., 
\Vampire\
\cite{RiazanovVoronkov:design_and_implementation_of_Vampire:AICOM:2002} or
\Gandalf\ \cite{Tammet:JAR:Gandalf:1997}.

\begin{sidewaysfigure} 
\psset{xunit=.1cm,yunit=.17cm,dash=5pt 4pt}
\begin{pspicture}(-100,2)(100,84)

\rput(-20,2){$answer(s_i)$}
\pnode(-20,4){u0}

\rput(-60,13){$;~query$}
\rput(-60,10){$\neg stud(X) \vee answer(X)$}
\pnode(-60,8){d1}

\rput(0,10){$stud(s_i)$}
\pnode(0,8){d2}
\pnode(0,12){u2}

\ncline{d1}{u0}
\ncline{d2}{u0}

\rput(-30,18){$pers(s_i)$}
\pnode(-30,16){d3}
\pnode(-30,20){u3}

\rput(20,18){$\neg pers(s_i) \vee stud(s_i)$}
\pnode(20,16){d4}
\pnode(20,20){u4}

\ncline{d3}{u2}
\ncline{d4}{u2}

\rput(-80,29){$;~grStud \sqsubseteq pers$}
\rput(-80,26){$\neg grStud(X) \vee pers(X)$}
\pnode(-80,24){d5}
\pnode(-80,28){u5}

\rput(-30,29){$;~DB~row$}
\rput(-30,26){$grStud(s_i)$}
\pnode(-30,24){d6}
\pnode(-30,28){u6}

\ncline{d5}{u3}
\ncline{d6}{u3}

\rput(-30,38){$\neg course(sk0(s_i)) \vee \neg pers(s_i) \vee stud(s_i)$}
\pnode(-30,36){d7}
\pnode(-20,40){u7}

\rput(60,26){$course(sk0(s_i))$}
\pnode(50,24){d8}
\pnode(50,28){u8}

\ncline{d7}{u4}
\ncline{d8}{u4}

\rput(50,48){$grCourse(sk0(s_i))$}
\pnode(50,46){d9}
\pnode(50,50){u9}

\rput(80,41){$;~grCourse \sqsubseteq course$}
\rput(80,38){$\neg grCourse(X) \vee course(X)$}
\pnode(80,36){d10}
\pnode(80,40){u10}

\ncline{d9}{u8}
\ncline{d10}{u8}

\rput(-65,51){$;~part~of~stud \equiv pers \sqcap \exists takesC.course$}
\rput(-65,48){$\neg takesC(X,Y) \vee \neg course(Y) \vee \neg pers(X) \vee stud(X)$}
\pnode(-65,46){d11}
\pnode(-65,50){u11}

\rput(-10,56){$takesC(s_i,sk0(s_i))$}
\pnode(-10,54){d12}
\pnode(-10,58){u12}

\ncline{d11}{u7}
\ncline{d12}{u7}

\rput(-60,69){$;~part~of~grStud \sqsubseteq \exists takesC.grCourse$}
\rput(-60,66){$\neg grStud(X) \vee takesC(X,sk0(X))$}
\pnode(-60,64){d13}
\pnode(-60,68){u13}

\rput(0,69){$;~DB~row$}
\rput(0,66){$grStud(s_i)$}
\pnode(0,64){d14}
\pnode(0,68){u14}

\ncline{d13}{u12}
\ncline{d14}{u12}

\rput(70,63){$;~DB~row$}
\rput(70,60){$grStud(s_i)$}
\pnode(70,58){d15}
\pnode(70,62){u15}

\rput(10,83){$;~part~of~grStud \sqsubseteq \exists takesC.grCourse$}
\rput(10,80){$\neg grStud(X) \vee grCourse(sk0(X))$}
\pnode(10,78){d16}
\pnode(10,82){u16}

\ncline{d15}{u9}
\ncline{d16}{u9}

\end{pspicture}
               \caption{Resoluton derivation of the answer $X := s_i$ for the query $stud(X)$.
                        \label{fig:ground_answer}}
\end{sidewaysfigure}
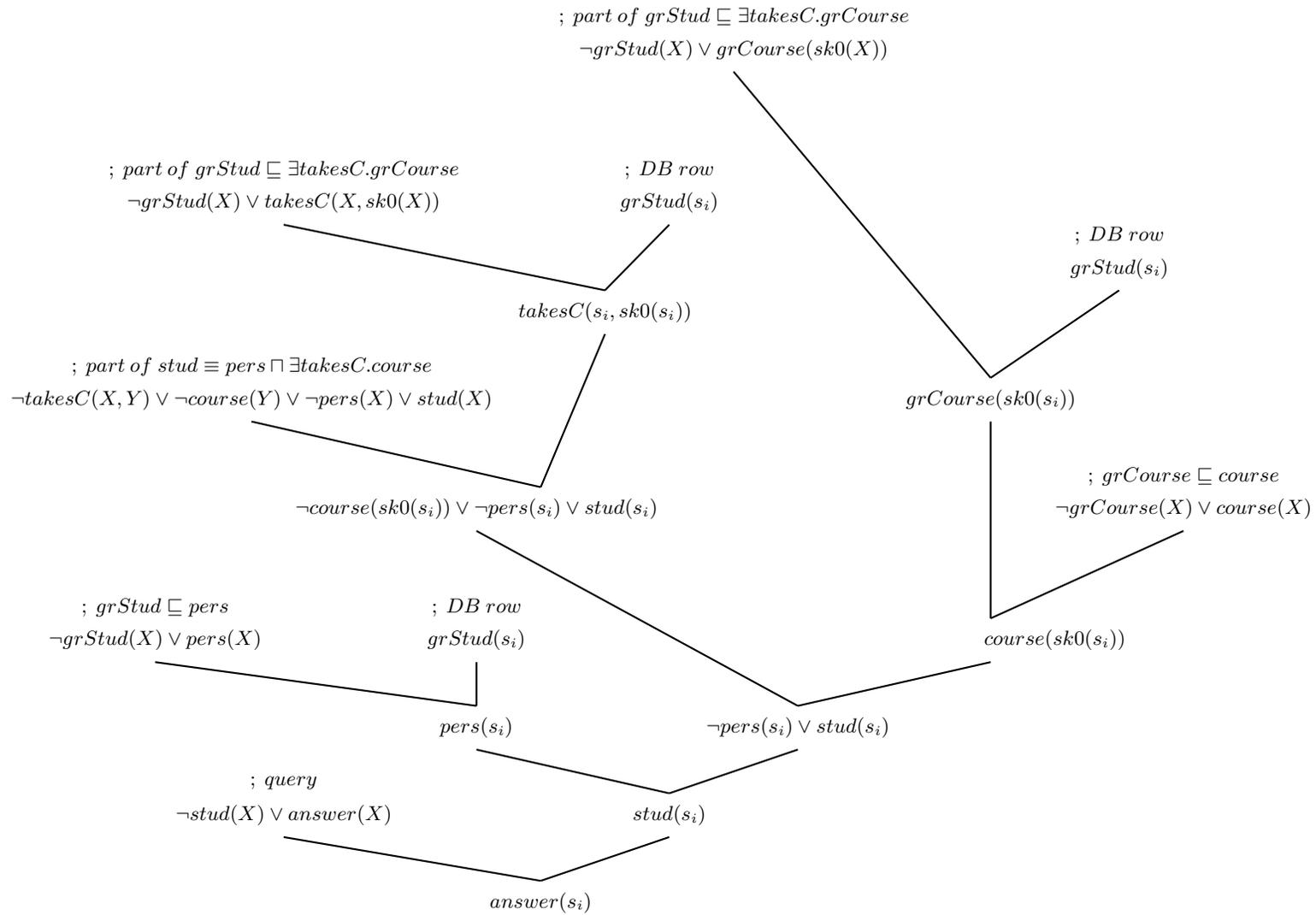

Even if we have enough memory to load the facts, this approach 
is likely to be very inefficient for the following reason.
If the RDB is large and the selectivity
of the query is not very high, we can expect that 
\emph{many answers will be obtained with structurally identical proofs}. 
For example, if our DB contains facts 
$graduateStudent(s_1), \ldots, graduateStudent(s_{100})$
(representing some table $graduateStudent$ which simply keeps a list of all
graduate students),
the facts will give rise to 100 answers to the query
$student(X)$\footnote{Query~6 from LUBM \cite{Guo+:ISWC:LUBM:2004}.},
each having a refutational proof of the form shown in 
Figure~\ref{fig:ground_answer} (where $grStud$, $takesC$,
$pers$ and $stud$ abbreviate $graduateStudent$, $takesCourse$,
$person$ and $student$, and $sk0$ is a Skolem function).

This example is intended to demonstrate how \emph{wasteful reasoning on the 
per-answer basis} is. Roughly speaking, the required amount of reasoning 
is multiplied with the number of answers. Even if the selectivity
of the query is very high, the reasoner is still likely to waste a lot of 
work in unsuccessful attempts represented by derivations not leading
to any answers.

Note that these observations are not too specific to the choice
of the reasoning method. For example, if we used \Prolog~or a tableaux-based
DL reasoner, we would have 
a similar picture: the same rule applications would be performed
for each answer $s_i$. 

\myparagraph{Main idea.} The main idea of our proposal is 
that \emph{answers with similar proofs
should be obtained in bulk}. More specifically, we propose to \emph{use
reasoning to find schematic answers} to queries, which can be 
later very efficiently \emph{instantiated by querying the RDB via 
the standard highly optimised RDBMS mechanisms}. 
Technically, we propose to search for the schematic answers
by \emph{reasoning on an abstraction of the RDB in some resolution-
and paramodulation-based calculus} 
(see \cite{BachmairGanzinger:HandbookAR:resolution:2001,Nieuwenhuis:HandbookAR:paramodulation:2001}). 
The abstraction 
and the reasoning on the abstraction should be organised in such 
a way that the obtained schematic answers can be turned into \emph{regular
RDBMS queries} (e.g., SQL queries).

\myparagraph{Constrained clauses and table abstractions.} To illustrate our main
idea, we apply it to the current example.
The clause $grStud(X)~|~grStud(X)$ is the \emph{abstraction} of the relevant
part of the RDB, i.e., it represents (generalises) all the facts 
$grStud(s_1),\ldots,grStud(s_{100})$. This is a very important feature
of our approach, so we emphasise that a potentially very large set of facts
is compactly represented with just one clause.
The part before ``$|$'' is the ordinary
logical content of the clause. What comes after ``$|$'' is a special constraint.
These constraints will be \emph{inherited} in all inference rules,
\emph{instantiated} with the corresponding unifiers and \emph{combined} 
when they come from different premises, just like, e.~g., ordering
or unifiability constraints in paramodulation-based theorem proving 
\cite{Nieuwenhuis:HandbookAR:paramodulation:2001}. 
Although our constraints
can be used as regular constraints -- that is to identify redundant inferences
by checking the satisfiability of the associated constraints 
w.r.t. the RDB (see Section~\ref{sec:constraints}) 
-- \emph{their main purpose
is to record which RDB fact abstractions contribute to a schematic answer
and what conditions on the variables of the abstractions have to be checked 
when the schematic answer is instantiated, so that the obtained concrete
answers are sound}. 

A derivation of a schematic answer for the 
query $student(X)$, covering all the concrete solutions 
$X := s_1,\ldots,X := s_{100}$, is shown in 
Figure~\ref{fig:general_solution_proof}. Note that the last inference simply
merges three identical atomic constraints. 
Also note that we write the answer literals
on the constraint sides of the clauses,
because they are not intended for resolution.

\begin{sidewaysfigure} 
\psset{xunit=.1cm,yunit=.16cm,dash=5pt 4pt}
\begin{pspicture}(-110,-6)(110,84)

\rput(-20,-6){$\emptyclause~|~\neg answer(X), grStud(X)$}
\pnode(-20,-4){u-1}

\rput(-20,2){$\emptyclause~|~\neg answer(X), grStud(X), grStud(X), grStud(X)$}
\pnode(-20,4){u0}
\pnode(-20,0){d0}

\ncline{d0}{u-1}

\rput(-60,13){$;~query$}
\rput(-60,10){$\neg stud(X)~|~\neg answer(X)$}
\pnode(-60,8){d1}

\rput(0,10){$stud(X)~|~grStud(X), grStud(X), grStud(X)$}
\pnode(0,8){d2}
\pnode(0,12){u2}

\ncline{d1}{u0}
\ncline{d2}{u0}

\rput(-30,18){$pers(X)~|~grStud(X) $}
\pnode(-30,16){d3}
\pnode(-30,20){u3}

\rput(20,18){$\neg pers(X) \vee stud(X)~|~grStud(X), grStud(X)$}
\pnode(20,16){d4}
\pnode(20,20){u4}

\ncline{d3}{u2}
\ncline{d4}{u2}

\rput(-80,29){$;~grStud \sqsubseteq pers$}
\rput(-80,26){$\neg grStud(X) \vee pers(X)$}
\pnode(-80,24){d5}
\pnode(-80,28){u5}

\rput(-30,29){$;~DB~table~abstraction$}
\rput(-30,26){$grStud(X)~|~grStud(X)$}
\pnode(-30,24){d6}
\pnode(-30,28){u6}

\ncline{d5}{u3}
\ncline{d6}{u3}

\rput(-30,38){$\neg course(sk0(X)) \vee \neg pers(X) \vee stud(X)~|~grStud(X)$}
\pnode(-30,36){d7}
\pnode(-20,40){u7}

\rput(60,26){$course(sk0(X))~|~grStud(X)$}
\pnode(50,24){d8}
\pnode(50,28){u8}

\ncline{d7}{u4}
\ncline{d8}{u4}

\rput(50,48){$grCourse(sk0(X))~|~grStud(X)$}
\pnode(50,46){d9}
\pnode(50,50){u9}

\rput(80,41){$;~grCourse \sqsubseteq course$}
\rput(80,38){$\neg grCourse(X) \vee course(X)$}
\pnode(80,36){d10}
\pnode(80,40){u10}

\ncline{d9}{u8}
\ncline{d10}{u8}

\rput(-65,51){$;~part~of~stud \equiv pers \sqcap \exists takesC.course$}
\rput(-65,48){$\neg takesC(X,Y) \vee \neg course(Y) \vee \neg pers(X) \vee stud(X)$}
\pnode(-65,46){d11}
\pnode(-65,50){u11}

\rput(-10,56){$takesC(X,sk0(X))~|~grStud(X)$}
\pnode(-10,54){d12}
\pnode(-10,58){u12}

\ncline{d11}{u7}
\ncline{d12}{u7}

\rput(-60,69){$;~part~of~grStud \sqsubseteq \exists takesC.grCourse$}
\rput(-60,66){$\neg grStud(X) \vee takesC(X,sk0(X))$}
\pnode(-60,64){d13}
\pnode(-60,68){u13}

\rput(0,69){$;~DB~table~abstraction$}
\rput(0,66){$grStud(X)~|~grStud(X)$}
\pnode(0,64){d14}
\pnode(0,68){u14}

\ncline{d13}{u12}
\ncline{d14}{u12}

\rput(70,63){$;~DB~table~abstraction$}
\rput(70,60){$grStud(X)~|~grStud(X)$}
\pnode(70,58){d15}
\pnode(70,62){u15}

\rput(10,83){$;~part~of~grStud \sqsubseteq \exists takesC.grCourse$}
\rput(10,80){$\neg grStud(X) \vee grCourse(sk0(X))$}
\pnode(10,78){d16}
\pnode(10,82){u16}

\ncline{d15}{u9}
\ncline{d16}{u9}

\end{pspicture}
               \caption{Resolution derivation of some schematic answer for $stud(X)$.
                        \label{fig:general_solution_proof}}
\end{sidewaysfigure}

\myparagraph{SQL generation.} Semantically the derived schematic answer
$\emptyclause~|~\neg answer(X), grStud(X)$ means 
that if some value $x$ is in the table $graduateStudent$, then $x$ is a legitimate
concrete answer to the query.
So, assuming that ${\bf id}$ is the (only) attribute in the RDB table
representing the instances of $graduateStudent$, the derived schematic answer
$\emptyclause~|~\neg answer(X), grStud(X)$ can be turned into 
the following simple SQL query:
\begin{center}
\begin{minipage}[t]{150mm}
\begin{tabbing}
SELECT {\bf id} AS $X$\\
FROM graduateStudent
\end{tabbing}
\end{minipage}
\end{center}
Evaluating this query over the RDB
will return all the answers $X := s_1,\ldots,X := s_{100}$.

\vspace{2mm}
Resolution reasoning on a DB abstraction may give rise 
to \emph{more than one schematic answer}. For example, 
$\emptyclause~|~\neg answer(X), grStud(X)$ does not necessarily cover all possible 
solutions of the initial query -- it only enumerates graduate students.
If our KB also postulates that any person taking a course is a student,
we want to select all such people as well.
So, suppose that our DB also contains the facts
$person(P_1),\ldots,person(P_{100})$, 
$takesCourse(P_1,C_1),\ldots,takesCourse(P_{100},C_{100})$ and 
$course(C_1),\ldots,course(C_{100})$ in the corresponding tables
$person$, $takesCourse$ and $course$.
These relations can be represented with the abstraction clauses
$person(X)~|~person(X)$,~~~$takesCourse(X,Y)~|~takesCourse(X,Y)$
and $course(X)~|~course(X)$.
Simple reasoning with these clauses modulo, say, a KB containing the rule
$student(P):-~person(P),~takesCourse(P,C),~course(C)$
or the DL axiom $person \sqcap \exists takesC.course \sqsubseteq student$,
produces the schematic
answer $\emptyclause~|~\neg answer(X), person(X), takesCourse(X,Y), course(Y)$.
Semantically it means that if table $takesCourse$ contains a record
$\{{\bf student}=s,{\bf course}=c\}$, and tables $person$ and $course$ contain $s$ and $c$
correspondingly, then $X := s$ is a legitimate concrete answer. 
Thus, the schematic answer can be turned into the following SQL query: 
\begin{center}
\begin{minipage}[t]{150mm}
\begin{tabbing}
SELECT person.{\bf id} AS $X$\\
FROM person, takesCourse, course\\ 
WHERE person.{\bf id} = takesCourse.{\bf student}\\ \dinc \inc
AND course.{\bf id} = takesCourse.{\bf course}
\end{tabbing}
\end{minipage}
\end{center}

The join conditions 
$person.{\bf id} = takesCourse.{\bf student}$
and 
$course.{\bf id} = takesCourse.{\bf course}$
reflect the fact that the corresponding arguments of the predicates
in the constraint attached to the schematic answer are equal:
e.g., the only argument of $person$, corresponding to $person.{\bf id}$,
and the first argument of $takesCourse$, corresponding to 
$takesCourse.{\bf student}$, are both the same variable $X$.

\myparagraph{Incremental query rewriting.} It bears repeating that,
in general, resolution over DB abstractions in the form of constrained clauses
may produce many, even infinitely many, schematic answers and, consequently,
SQL queries. They are produced
one by one, and the union of their answers covers the whole set of concrete 
answers to the query.
If there is only a finite number of concrete answers, e.~g., if the query 
allows concrete answers to contain only plain data items from the database, 
then all concrete answers are covered after some finite number of steps.
In a sense, the original semantic query is rewriten as a sequence of SQL queries, 
so we call our technique \emph{incremental query rewriting}.

\myparagraph{Benefits.} The main advantage of the proposed scheme is 
the \emph{expressivity scalability}. 
For example, in applications not requiring termination, the expressivity
of the knowledge representation formalisms is only limited by the expressivity
of the full FOL\footnote{Complete methods for efficient schematic
answer finding in FOL \emph{with equality} are yet to be formulated and proved fomally 
(see the brief discussion in Section~\ref{sec:future_work}).}, 
although specialised treatment 
of various FOL fragments is likely to be essential for good performance. 
The use of such a powerful logic as FOL as the common platform 
also allows easy practical simultaneous use of heterogeneous knowledge bases,
at least for some data retrieval tasks. In particular, it means that 
users can freely mix all kinds of OWL and RDFS ontologies with all kinds of 
(monotonic) declarative rule sets, e.~g., in RuleML or SWRL.

It is important that we don't pay too high a price in terms of performance,
for the extra expressivity. The method has good data scalability: 
roughly, \emph{the cost of reasoning is not multiplied by the volume
of data}. Note also that we don't have to do any static conversion of the data
into a different data model, e.~g., RDF triples or OWL ABox
-- querying can be done on live databases via the hosting RDBMSs. 
All this makes our method potentially usable with very large 
databases in real-life settings.

An additional advantage of our approach is that answers to semantic queries
can be relatively easily given rigorous explanations. Roughly speaking,
if we need to explain a concrete answer, we simply instantiate the derivation
of the corresponding schematic answer by replacing DB table abstractions with
concrete DB rows, and propagating this data through the derivation. 
Thus, we obtain a resolution proof of the answer, which can be relatively
easily analysed or transformed into a more intuitive representation.


   \section{Soundness and completness of schematic answer computation.}
       \label{sec:soundness_and_completeness}


So far we have only speculated that schematic answer search
can be implemented based on resolution. In this section we are 
going to put it on a formal basis. We will show that in the context
of FOL without equality some popular resolution-based methods
can deliver the desired results. In particular, we will characterise
a class of resolution-based calculi that are both sound and complete
for query answering over database abstractions.

We assume familiarity of the reader with the standard notions of
first-order logic, such as terms, formulas, literals and clauses,
substitutions, etc., and some key results, such as the Herbrand's theorem.
Bibliographic references are provided for more specialised concepts
and facts.

\subsection{Definitions.}

\myparagraph{Deductive queries.} In our settings, a \emph{deductive query}
is a triple $\langle DB,KB,\varphi \rangle$, where 
(i) the logical representation $DB$ of some relational database is a set
of ground atomic non-equality formulas, each representing a row in a table
in the database, 
(ii) the \emph{knowledge base} $KB$ is a finite set of FOL axioms,
corresponding to both the domain ontologies and semantic RDB schema mappings 
in our scenario,
and (iii) the \emph{goal} $\varphi$ of the query is a construct of the 
form $\langle X_1,\ldots,X_k \rangle \langle Y_1,\ldots,Y_m\rangle C$,
where $C$ is a nonempty clause,
$k,m \geq 0$, $\{X_1,\ldots,X_k,Y_1,\ldots,Y_m\} = vars(C)$,
all $X_i$ and $Y_i$ are pairwise distinct.
We call $X_i$ \emph{distinguished variables}, 
and $Y_j$ \emph{undistinguished variables} of the query.
Intutively, the deductive query represents a request to find
all $X_i$, such that there exist some $Y_j$, such that 
$\varphi(\overline{X},\overline{Y})$ is \emph{inconsistent} with
$DB \cup KB$. In other words, answers to the query refute $\varphi$
rather than prove it. This convention is made for technical convenience. 
Users of our technology can work in terms of positive queries.

\myparagraph{Recording literals.} 
In our settings, a clause with \emph{recording literals}\footnote{
We prefer this to the more general term ``constrained clause''
because we want to emphasise the nature and the role of our constraints,
and to avoid confusion with other kinds of constraints used in automated reasoning
and logic programming. 
} is a construct of the following form: $C~|~\gamma$, where $C$ 
is a regular first-order clause, possibly empty, and
$\gamma$ is a finite multiset of literals, possibly empty.
We will say that the literals of $\gamma$ are \emph{recording literals}. 

\emph{Semantically},
$C~|~\lambda_1, \ldots , \lambda_n$ is the same as 
the regular clause 
$C \vee \overline{\lambda_1} \vee \ldots \vee \overline{\lambda_n}$, which will
be denoted as $Sem(C~|~\lambda_1, \ldots , \lambda_n)$.
All semantic relations between $Sem(C~|~\gamma)$ and other formulas
are transfered to $C~|~\gamma$. For example, when we say that  $C~|~\gamma$
is implied by something, it means that $Sem(C~|~\gamma)$ is implied,
and vice versa.

Regular clauses will be often identified with clauses with empty recording
parts, i.e., we will not distinguish $C$ from $C~|~\emptyset$.

We say that a clause $C'~|~\gamma'$ subsumes the clause $C~|~\gamma$
iff there is a substitution $\theta$ that makes $C'\theta$ a submultiset
of $C$, and $\gamma'\theta$ a submultiset of $\gamma'$. In this case
we will also say that $C'~|~\gamma'$ is a \emph{generalisation} of $C~|~\gamma$.

\myparagraph{Concrete and schematic answers.}
We distinguish a special predicate symbol $@$\footnote{Corresponds to the predicate 
$answer$ used in our previous examples.}. 
A ground atomic formula $@(t_1,\ldots,t_k)$ 
is a \emph{concrete} answer to the deductive query
$\langle DB,KB, \langle X_1,\ldots,X_k \rangle \langle Y_1,\ldots,Y_m\rangle C\rangle$,
 if the clause
$C[X_1/t_1,\ldots,X_k/t_k]$ is \emph{inconsistent} with $DB \cup KB$
or, equivalently, the formula $\exists Y_1 \ldots Y_m \neg C[X_1/t_1,\ldots,X_k/t_k]$
is implied by $DB \cup KB$.

We say that a clause $\emptyclause~|~\gamma$ is a \emph{schematic answer} to 
a deductive query
$\langle DB,KB, \langle X_1,\ldots,X_k \rangle \langle Y_1,\ldots,Y_m\rangle C\rangle$, 
if every atomic ground formula of the form $@(t_1,\ldots,t_k)$ implied by 
$DB \cup \{ \emptyclause~|~\gamma \}$, is a concrete answer to the query. Every such concrete
answer will be called an \emph{instance} of the schematic answer.

\myparagraph{Database abstractions.}
In our settings, a finite set $DB'$ of 
clauses of the form $p(t_1,\ldots,t_k)~|~p(t_1,\ldots,t_k)$ is 
an \emph{abstraction} of the logical representation $DB$ of a database
if for every atomic formula $\rho \in DB$, there is a clause 
$\rho'~|~\rho' \in DB'$ and a substitution $\theta$, such that 
$\rho'\theta = \rho$.
Note that \emph{semantically} all clauses in $DB'$ are tautologies, because
$Sem(p(t_1,\ldots,t_k)~|~p(t_1,\ldots,t_k)) ~~=~~ p(t_1,\ldots,t_k) \vee \neg p(t_1,\ldots,t_k)$.

The simplest kind of an abstraction for an RDB is the set of all clauses
$p(X_1,\ldots,X_k)~|~p(X_1,\ldots,X_k)$, where all $X_i$ are pairwise distinct
variables, and each $p$ corresponds to a table in the RDB. Dealing with such an abstraction
can be viewed as reasoning on the schema of the RDB. However, in principle, we
can have more specific abstractions. For example, if we know that the first column
of our RDB table $p$ contains only values $a$ and $b$, we may choose to have two abstraction
clauses: $p(a,X_2,\ldots,X_k)~|~p(a,X_2,\ldots,X_k)$ and
$p(b,X_2,\ldots,X_k)~|~p(b,X_2,\ldots,X_k)$\footnote{Moreover, we can have just one
abstraction clause, e.~g.,  $p(X_1,\ldots,X_k)~|~p(X_1,\ldots,X_k),~X_1 \in \{a,b\}$
with the additional \emph{ad hoc constraint} $X_1 \in \{a,b\}$, 
but this kind of optimisations is outside the scope of this research note.
}.

\myparagraph{Calculi.}
In this research note we only deal with calculi that are sound and 
complete variants of resolution\footnote{Paramodulation is also briefly
discussed as a future research opportunity in Section~\ref{sec:future_work}.}
(see, e.~g., \cite{BachmairGanzinger:HandbookAR:resolution:2001}).
All inference rules in these calculi are of the form 
\begin{center}          
 \[
      \infer
       {D}
       {C_1 ~~~  C_2 ~~ \ldots ~~ C_n }
    \]
\end{center} 
where $C_i$ and $D$ are ordinary clauses, and $n \geq 1$. Most such rules
have a substitution $\theta$ associated with them, which is required 
to unify some subexpressions in $C_i$, usually atoms of complementary literals.
Rules in the calculi that are
of interest to us can be easily extended to clauses with recording
literals:
\begin{center}          
 \[
      \infer
       {D~|~\gamma_1\theta, \gamma_2\theta, \ldots, \gamma_n\theta }
       {C_1~|~\gamma_1 ~~~  C_2~|~\gamma_2 ~~ \ldots ~~ C_n~|~\gamma_n }
    \]
\end{center} 

So, for example, here is the binary resolution rule extended to clauses with recording
literals:
\begin{center}          
 \[
      \infer
       {C'_1\theta \vee C'_2\theta~|~\gamma_1\theta, \gamma_2\theta}
       {C'_1 \vee A~|~\gamma_1 ~~~~~~  C'_2 \vee \neg B~|~\gamma_2}
    \]
\end{center} 
where $\theta$ is the most general unifier of the atoms $A$ and $B$.

If a calculus $R'$ is obtained by extending the rules of a calculus $R$
to clauses with recording literals, we will simply say that $R'$ is 
a \emph{calculus with recording literals} and $R$ is its 
\emph{projection to regular clauses}.

Apart from nonredundant inferences,
resolution calculi used in practice usually include some \emph{admissible}
redundant inferences. Implementors have the freedom of performing or not
performing such inferences without affecting the completeness of the reasoning
process. However, for the purposes of this research note it is convinient to assume that
calculi being considered only contain nonredundant inferences. This assumption
does not affect generality.

A calculus with recording literals is \emph{sound} if 
$Sem$ of the conclusion of every derivation is logically implied by 
the $Sem$ images of the clauses in the leaves. 
It is obvious that a calculus with recording literals is sound if its
projection to regular clauses is sound because recording literals are fully
inherited. 
A calculus with recording literals
is \emph{refutationally complete} if its projection to regular clauses is 
refutationally complete, i.e., an empty clause can be derived from 
any unsatisfiable set of clauses.

In this research note we will mention \emph{fully specified calculi} to distinguish 
them from generic (parameterised) calculi. For example, 
the ordered binary resolution in general is not fully specified -- it is a
generic calculus \emph{parameterised} by an order on literals. 
If we fix this
parameter by specifying a concrete order, we obtain a fully specified
calculus. We view a fully specified calculus as the set of all its 
elementary inferences.

We say that a fully specified calculus $R$ with recording literals
is \emph{generalisation-tolerant} if every inference in $R$ is
generalisation-tolerant. An elementary inference 
\begin{center}          
 \[
      \infer
       {D~|~\delta}
       {C_1~|~\gamma_1 ~~~  C_2~|~\gamma_2 ~~ \ldots ~~ C_n~|~\gamma_n}
    \]
\end{center} 
from the calculus R is generalisation-tolerant if
for every generalisation $C'_i~|~\gamma'_i$ of a premise $C_i~|~\gamma_i$,
the calculus $R$ also contains an elementary inference of some generalisation
$D'~|~\delta'$ of $D~|~\delta$, where the premises are a submultiset
of $\{C_1~|~\gamma_1,~ \ldots,~ C_{i-1}~|~\gamma_{i-1},~~~ C'_i~|~\gamma'_i,~~~ C_{i+1}~|~\gamma_{i+1},~ \ldots,~ C_n~|~\gamma_n\}$.

Unordered binary resolution and hyperresolution provide simple 
examples of generalisation-tolerant calculi. Their ordered versions 
using admissible orderings (see, e.~g., 
\cite{BachmairGanzinger:HandbookAR:resolution:2001})
also cause no problems because application of generalisation 
to a clause
cannot make a maximal literal nonmaximal, because of the 
\emph{substitution property} of admissible orderings: $L_1 > L_2$ implies 
$L_1\theta > L_2\theta$.
Adding (negative) literal selection (see, e.~g., 
\cite{BachmairGanzinger:HandbookAR:resolution:2001})
requires some care. In general, if a literal is selected in a clause,
its image, if it exists, in any generalisation should be selected too. 
Such selection functions are still possible. For example,
we can select \emph{all} negative literals that are maximal 
w.~r.~t. some ordering satisfying the substitution property.
In this case, however, we can no longer restrict ourselves 
to selecting a single literal in a clause, because the ordering
can only be partial. 

Note that such calculi are the main working horses in several
efficient FOL reasoners, e.~g., \Vampire.

\subsection{Soundness.}

\myparagraph{Theorem 1.} Suppose $R$ is a sound fully specified calculus 
with recording literals. Consider a deductive query 
$Q = \langle DB,KB, \langle X_1,\ldots,X_k \rangle \langle Y_1,\ldots,Y_m \rangle C \rangle$.
Suppose $DB'$ is an abstraction of $DB$. Suppose we can derive in $R$
a clause $\emptyclause~|~\gamma$ from 
$DB' \cup KB \cup \{ C~|~\neg @(X_1,\ldots,X_k) \}$.
Then $\emptyclause~|~\gamma$ is a schematic answer to $Q$.

\myparagraph{Proof.} Suppose $DB \cup \{ \emptyclause~|~\gamma \}$ implies a ground formula $@(t_1,\ldots,t_k)$. 
We have to show that $@(t_1,\ldots,t_k)$ is a concrete answer to $Q$, 
i.e., $DB \cup KB \cup \{C[X_1/t_1,\ldots,X_k/t_k]\}$ is unsatisfiable. 

Since $R$ is sound, $\emptyclause~|~\gamma$ is derived from 
$DB' \cup KB \cup \{ C~|~\neg @(X_1,\ldots,X_k) \}$ and 
$DB'$ contains only clauses that are semantically tautologies, 
the clause $\emptyclause~|~\gamma$ 
is implied by $KB \cup \{C~|~\neg @(X_1,\ldots,X_k)\}$. 
Under our assumption, this means that 
$DB \cup KB \cup \{ C~|~\neg @(X_1,\ldots,X_k) \}$ implies $@(t_1,\ldots,t_k)$. 
Note that the predicate $@$ does not occur in $DB$, $KB$ or $C$ and, 
therefore, $DB \cup KB \cup \{ C[X_1/t_1,\ldots,X_k/t_k]\}$ is unsatisfiable.

\subsection{Completeness.}

\myparagraph{Theorem 2.} Suppose $R$ is a refutationally complete 
and generalisation-tolerant fully specified calculus
with recording literals. Consider a deductive query 
$Q = \langle DB,KB, \langle X_1,\ldots,X_k \rangle \langle Y_1,\ldots,Y_m
\rangle C \rangle$. Suppose $DB'$ is an abstraction of $DB$.
Then, for every concrete answer $@(t_1,\ldots,t_k)$ to $Q$
one can derive in $R$ from 
$DB' \cup KB \cup \{ C~|~\neg @(X_1,\ldots,X_k) \}$ a clause
$\emptyclause~|~\gamma$, such that $@(t_1,\ldots,t_k)$
is an instance of the schematic answer $\emptyclause~|~\gamma$.

\myparagraph{Proof.} The refutational completeness of $R$ means
that we can construct a refutation $\Delta$ of 
$DB \cup KB \cup C[X_1/t_1,\ldots,X_k/t_k]$. The main idea of this
proof is that in a generalisation-tolerant calculus finding an answer
to a query is not much more difficult than just proving the answer. 
Technically, we will convert $\Delta$ into a derivation of a schematic answer
covering the concrete answer $@(t_1,\ldots,t_k)$.

Assume that $\rho_i$, $i \in [1 \ldots p]$, are all the facts from 
$DB$ that contribute to $\Delta$ (as leaves of the refutation).
We can convert $\Delta$ into a derivation $\Delta'$ of 
a clause of the form 
$\emptyclause~|~\rho_1, \ldots, \rho_p, 
\neg A_1, \ldots, \neg A_n$, where $p,n \geq 0$ and all atoms 
$A_i = @(t_1,\ldots,t_k)$,
from the clauses 
$\rho_1~|~\rho_1, ~~\ldots,~~\rho_p~|~\rho_m$,
$C[X_1/t_1,\ldots,X_k/t_k]~|~\neg @(t_1,\ldots,t_k)$ and some clauses from $KB$.
To this end, we simply add the recording literals in the corresponding
leaves of $\Delta$ and propagate them all the way to the root.
Obviously, 
$DB \cup \{\emptyclause~|~\rho_1,\ldots,\rho_m, \neg A_1, \ldots, \neg A_n\}$
implies $@(t_1,\ldots,t_k)$.

To complete the proof, we will show that 
$\Delta'$ can be converted into a derivation of a generalisation 
$\emptyclause~|~\gamma$
for the clause 
$\emptyclause~|~\rho_1,\ldots,\rho_m, \neg A_1, \ldots, \neg A_n$
from $DB' \cup KB \cup \{ C~|~\neg @(X_1,\ldots,X_k) \}$. 
This is a corollary of a more general statement: if we can derive 
some clause $D$ from clauses $C_1,\ldots,C_q$ in $R$, and
$C'_1,\ldots,C'_q$ are some generalisations of those clauses, 
then there is a derivation from some of $C'_1,\ldots,C'_q$ 
in $R$ of some generalisation $D'$ of $D$. 
This can be easily proved by induction on
the complexity of the derivation, taking into account 
the generalisation-tolerance of $R$. 

Finally, note that $\emptyclause~|~\gamma$ 
implies $\emptyclause~|~\rho_1,\ldots,\rho_m, \neg A_1, \ldots, \neg A_n$, 
and therefore $DB \cup \{\emptyclause~|~\gamma\}$ implies $@(t_1,\ldots,t_k)$.


                 \section{Recording literals as search space pruning constraints.}
                    \label{sec:constraints}

Let us make an important observation: \emph{some schematic answers to 
deductive queries cover no concrete answers}. These schematic answers
are useless and the work spent on their generation is wasted. 
We can address this problem by trying to block search directions
that can only lead to such useless schematic answers.

Suppose we are searching for schematic answers to 
$\langle DB,KB, \langle X_1,\ldots,X_k \rangle \langle Y_1,\ldots,Y_m
\rangle C \rangle$ by deriving consequences of 
$DB' \cup KB \cup \{ C~|~\neg @(X_1,\ldots,X_k) \}$ in an appropriate calculus,
where $DB'$ is an abstraction of $DB$. 

\subsection{Database abstraction literals.}
  \label{subsec:constraints:db_abstraction_literals} 
Suppose we have derived a clause  
$E = ~~D~|~\rho'_1, \ldots, \rho'_p, \neg A_1, \ldots, \neg A_n$
where $p > 0, n \geq 0$, all the atoms $A_i$ are of the form
$@(t^i_1,\ldots,t^i_k)$ and all the literals $\rho'_j$ are 
inherited from the recording literals of clauses from $DB'$. 
We can treat $\rho'_1, \ldots, \rho'_p$ as follows: 
if we can somehow establish that 
the constraint $\rho'_1, \ldots, \rho'_p$ \emph{has no solutions} w.~r.~t. $DB$,
we can remove the clause $E$ from the search space. 
A \emph{solution} of $\rho'_1, \ldots, \rho'_p$ w.~r.~t. $DB$ is a substitution
$\theta$, such that all $\rho'_i\theta \in DB$.

Such a treatment can be justified with the following argument.
It is obvious that if
$\rho'_1, \ldots, \rho'_p$ has no solutions w.~r.~t. $DB$,
then any more specific constraint
$\rho'_1\sigma, \ldots, \rho'_p\sigma$, where $\sigma$
is some substitution, also has no solutions. 
Since all recording literals are fully inherited in the calculi we are 
dealing with, any clause derived from 
$E$ and any other clauses, will have the same property. Therefore,
any schematic answer $\emptyclause~|~\gamma$ whose derivation contains
the clause, will contain in $\gamma$ a nonempty subconstraint without $@$, 
having no solutions w.~r.~t. $DB$. Thus, $\emptyclause~|~\gamma$ cannot cover
any concrete answers because the non-$@$ part of the constraint $\gamma$
cannot be satisfied. 

To summarise, we can discard clauses like $E$
without sacrificing the completeness w.~r.~t. concrete answers.
Practically, this can be done by converting 
$\rho'_1, \ldots, \rho'_p$ into an SQL query
(similar to how it is done in Section~\ref{sec:sql-generation} 
for schematic answers) and evaluating the query on the database
-- empty result set indicates absense of solutions w.~r.~t. $DB$.

\subsection{Answer literals.}
  \label{subsec:constraints:answer_literals} 
Suppose we have derived a schematic answer 
$\emptyclause~|~D, \neg A_1, \ldots, \neg A_n$
where $D$ only contains database abstraction literals or is empty,
and $n > 0$. For the schematic answer to have instances,
the answer literals $\neg A_i$ must be simultaneously unifiable. 
Indeed, suppose $@(t_1,\ldots,t_k)$ is an instance of the schematic answer. 
By Herbrand's theorem,
$DB \cup \{ \neg @(t_1,\ldots,t_k) \}$ is inconsistent with 
a finite set of ground clauses of the form 
$\emptyclause~|~D\theta, \neg A_1\theta, \ldots, \neg A_n\theta$.
We assume that the set is minimal. It cannot be empty because $@$ does not
occur in $DB$ and $DB$ itself is trivially consistent.
Consider any clause $\emptyclause~|~D\theta, \neg A_1\theta, \ldots, \neg A_n\theta$
from the set. All the atoms $A_i\theta$ from this clause are equal
to $@(t_1,\ldots,t_k)$ because otherwise the set would not be minimal
-- any model of the set without this clause could be extended to make this
clause true by making an appropriate $A_i\theta$ true. 
Thus, all $A_i$ are simultaneously unifiable.

The fact proved above can be used to prune the search space as follows:
if we derive an intermediate clause with some $@$-literals that are not 
simultaneously unifiable, we can discard the clause because any schematic
answer derived from it will have no instances. 
Moreover, we can use the most general unifier for $@$-literals 
to strengthen the test on database abstraction literals
by applying the unifier to them before solving them on 
the database.


                 \section{SQL generation.}
                    \label{sec:sql-generation}


Suppose that we have found a schematic answer
$\emptyclause~|~\rho_1, \ldots, \rho_p, \neg A_1, \ldots, \neg A_n$ 
to a query 
$\langle DB,KB, \langle X_1,\ldots,X_k \rangle \langle Y_1,\ldots,Y_m
\rangle C \rangle$.
Now our task is to enumerate all instances of the schematic answer
by querying the relational database modeled by the fact set $DB$, 
with an SQL query.

We have four cases to consider. 
(1) If $p = n = 0$, then we simply have
a refutation of $KB$. Formally, this means that any 
ground $@(t_1,\ldots,t_k)$ is a correct answer, but for practical purposes 
this is useless. Instead, we should simply inform the user about 
the inconsistency. 
(2) If $p = 0$ but $n \neq 0$, we have to try 
to unify all the literals $A_i$. If $\theta = mgu(A_1,\ldots,A_n)$,
then the set of instances of the schematic answer coincides with 
the set of ground instances of $A_1\theta$.  
(3) If $p \neq 0$ but $n = 0$, there is a possibility 
that $DB \cup KB$ is inconsistent. We may want to check this possibility
by checking if $\rho_1, \ldots, \rho_p$ has solutions over $DB$
-- if it does, $DB$ is inconsistent with $KB$. The check itself 
can be done by converting 
$\rho_1, \ldots, \rho_p$ into an SQL query as in 
the next case, and checking if an answer to the SQL query exists.
(4) In the rest of this section we will be considering the most interesting
case when $p \neq 0$ and $n \neq 0$.

\subsection{Merging and flattening answer literals.}

In fact, we only need to consider the case when $n = 1$. 
Indeed, if $\emptyclause~|~\gamma$, where 
$\gamma = D, \neg A_1, \ldots, \neg A_n$ is a schematic answer, 
it is only interesting to us if all $A_i$ are simultaneously unifiable,
as demonstrated in Section~\ref{sec:constraints}. So, suppose 
$\theta = mgu(A_1,\ldots,A_n)$. We are going to show that 
$\emptyclause~|~\gamma$ and $\emptyclause~|~\gamma\theta$
cover the same sets of concrete answers. Suppose that $@(t_1,\ldots,t_k)$ is an
instance of $\emptyclause~|~\gamma$.
Keeping in mind the Herbrand's theorem, 
consider a minimal set of ground clauses of the form
$\emptyclause~|~\gamma\theta_j$, $j = 1 \ldots m$, inconsistent with 
$DB \cup \{ \neg @(t_1,\ldots,t_k) \}$. As was shown in 
Section~\ref{subsec:constraints:answer_literals}, all $@$-literals in  
the clauses $\emptyclause~|~\gamma\theta_j$ are equal to 
$\neg @(t_1,\ldots,t_k)$.
Assuming that $\theta$ only affects variables occuring in $A_i$,
i.~e., all other variables in $\gamma$ are only renamed by $\theta$
(otherwise the unifier $\theta$ would not be the most general one),
we can conclude that $\emptyclause~|~\gamma\theta$ subsumes each of the clauses 
$\emptyclause~|~\gamma\theta_j$ and, 
therefore, $DB \cup \{ \neg @(t_1,\ldots,t_k) \} \cup \{ \emptyclause~|~\gamma\theta \}$
is unsatisfiable, which makes $@(t_1,\ldots,t_k)$ an instance of 
$\emptyclause~|~\gamma\theta$. The opposite direction is trivial. 
Now, since we can replace $\gamma$ with
$\gamma\theta$, we can also replace it with the shorter
constraint $D\theta, \neg A_1\theta$.

We can make another simplifying assumption: we only have to deal with
schematic answers of the form $\emptyclause~|~D, \neg @(X_1,\ldots,X_k)$, 
where $X_i$ are
pairwise distinct variables, each $X_i$ occurs in $D$, and $D$ contains 
only database abstraction
literals. If we need to enumerate instances of $\emptyclause~|~D, \neg A$ 
where $A$ is a more complex
$@$-literal, we enumerate instances of 
$\emptyclause~|~D, \neg @(X_1,\ldots,X_k)$,
where $\{X_1,\ldots,X_k\} = vars(A) \cap vars(D)$, and for every such instance 
$@(t_1,\ldots,t_k)$ we construct all instances
$A[X_1/t_1,\ldots,X_k/t_k]\sigma$ of the original schematic answer, 
where the substitutions $\sigma$ can instantiate variables not occuring 
in $D$ with arbitrary ground terms. 
In practice, we can explicitly report such variables as universally quantified 
in the answers, which is the approach adopted by most, if not all, Prolog
implementations.

\subsection{Flattening the database abstraction literals.}

Recall that all facts in $DB$ are of the form
$r_i(a^i_1,\ldots)$, where the predicates $r_i$ correspond 
to tables in a relational database and all $a^i_j$ are constants.
This and the considerations from
Section~\ref{subsec:constraints:db_abstraction_literals} 
justify the following assumption: literals from $D$ do not contain
compound terms, i.~e., all their arguments are variables or constants.
If this condition is false, the schematic answer is simply useless 
because $D$ has no solutions w.~r.~t. $DB$.

One final transformation of schematic answers is needed to make the SQL
query generation straightforward. Namely, we can represent the schematic
answer with a \emph{semantically equivalent} clause of the form
$E_a \vee E_c \vee E_d \vee D_x \vee A$, where 
(i) $A = @(X_1,\ldots,X_k)$ and all \emph{answer variables} 
$X_i$ are pairwise distinct;
(ii) $D_x = \neg r_1(Y^1_1,\ldots,Y^1_{k(1)}) 
          \vee \ldots \vee
          \neg r_p(Y^p_1,\ldots,Y^p_{k(p)})$ and all variables
$Y^i_j$ are pairwise distinct;
(iii) $E_a$ consists of $k$ negative equality literals
$\alpha_i \not \simeq X_i$, $i = 1 \ldots k$,
where $\alpha_i \in \{Y^1_1,\ldots,Y^p_{k(p)}\}$;
(iv) $E_c$ consists of zero or more negative equality literals
of the form $\alpha \not \simeq \beta$, where 
$\alpha \in \{Y^1_1,\ldots,Y^p_{k(p)}\}$ and $\beta$ is a constant;
(v) $E_d$ consists of zero or more negative equality literals
of the form $\alpha \not \simeq \beta$, where 
$\alpha,\beta \in \{Y^1_1,\ldots,Y^p_{k(p)}\}$.

Here is a sketch of an algorithm for the transformation.
Suppose we initially have 

$
\neg r_1(t^1_1,\ldots,t^1_{k(1)}) 
 \vee \ldots \vee
 \neg r_p(t^p_1,\ldots,t^p_{k(p)})
 \vee
 @(X_1,\ldots,X_k)$, 
which is just 
$Sem(\emptyclause~|~r_1(t^1_1,\ldots,t^1_{k(1)}), 
     \ldots, 
     r_p(t^p_1,\ldots,t^p_{k(p)}),
     \neg @(X_1,\ldots,X_k)) 
$. 
Recall that all $t^i_j$ are variables or constants.
We transform it into the equivalent clause 
$E 
   \vee 
 \neg r_1(Y^1_1,\ldots,Y^1_{k(1)}) 
    \vee \ldots \vee
   \neg r_p(Y^p_1,\ldots,Y^p_{k(p)})
 \vee
 @(X_1,\ldots,X_k)$
where $E$ consists of all literals of the form 
$Y^i_j \not \simeq t^i_j$. 
Now $E_c$ can be easily extracted from $E$ by taking all 
$Y^i_j \not \simeq t^i_j$ where $t^i_j$ are constants.
$E_a$ is obtained by taking one literal of the form 
$Y^i_j \not \simeq X_e$ for each answer variable $X_e$. 
Note that we have some nondeterminism here because
the same answer variable can occur in more than one literal in $E$.
Finally, $E_d$ is obtained by computing
the set $\{ Y^i_j \not \simeq Y^u_v |~ 
    for~some~nonconstant~t,~~t \not \simeq Y^i_j, t \not \simeq Y^u_v \in E\}$
and removing redundant literals from it. In this context, 
a literal $Y^i_j \not \simeq Y^u_v$ is redundant if there are 
two literals $Y^i_j \not \simeq Y^r_s$ and $Y^r_s \not \simeq Y^u_v$
(modulo the symmetry of the equality predicate). Note that this process 
is also nondeterministic because the result depends on the order
of removing redundant literals.

\subsection{Forming the SQL query.}

The transformed schematic answer $E_a \vee E_c \vee E_d \vee D_x \vee A$
can be translated into an SQL query of the form
$SELECT~~\langle columns \rangle~~FROM~~\langle tables \rangle~~WHERE~~\langle join~conditions \rangle$.

The expression $\langle columns \rangle$ is a comma-separated 
list of answer column declarations of the form 
$R_i.\#_j~~AS~~X_e$ for each $Y^i_j \not \simeq X_e \in E_a$.
Here $R_i$ is a fresh table alias corresponing to the literal
$r_i(Y^i_1,\ldots,Y^i_{k(i)})$ and $\#_j$ denotes the $j$-th attribute
name in the table $r_i$ from our RDB schema. 

The expression $\langle tables \rangle$ is a comma-separated
list of table aliases of the form $r_i~~AS~~R_i$, $i = 1 \ldots p$.
We have to use aliases because, in general, some $r_i$ may coincide. 

The expression $\langle join~conditions \rangle$ is a conjunction
of elementary join condition of two kinds:
(i) $R_i.\#_j = \beta$ for each $Y^i_j \not \simeq \beta \in E_c$,
and (ii) $R_i.\#_j = R_u.\#_v$ for each $Y^i_j \not \simeq Y^u_v \in E_d$.


                 \section{A note on indexing Semantic Web documents with data abstractions.}
                    \label{sec:indexing_with_abstractions}


In the context of Semantic Web (SW), it is important to be able
to index distributed semantic data description sets (SW~documents, for simplicity),
so that, given a semantic query modulo some knowledge bases, we can load only the 
SW~documents that are potentially relevant to the query. In this section
we briefly sketch a possible scheme for such indexing that is compatible
with our approach to deductive querying. 

Conventional search engines index regular Web documents by words appearing in them. 
We cannot simply follow this example by indexing SW~documents 
by the names of objects, concepts and
relations occuring in them. This is so because retrieval in general 
may require reasoning, and thus the relevant documents may use no common 
symbols with the query. For example, a query may request to find animals
of bright colours. If some SW~document describes, e.~g., pink elephants, it is relevant,
but lexically there is no overlap with the query. Only reasoning reveals
the relation between ``http://zooontology.org/concept\#elephant'' and 
``http://zooontology.org/concept\#animal'', and between 
``http://www.colors.org/concept\#pink'' and 
``http://www.colors.org/concept\#bright\_colour''. 

Note that conceptually there is hardly any difference between RDBs
and, say, OWL data description sets based on the Web: an RDB can be 
\emph{modeled} as a set of ground atomic logical assertions, 
and, practically, an SW~document \emph{is} such a set. So, just like we use
abstractions to represent relational data compactly in reasoning,
we can use abstractions to represent SW~documents.
For example, a potentially large SW~document introducing many
pink elephants can be compactly represented by its abstraction
$zoo{:}elephant(X)~|~zoo{:}elephant(X)$,
$colors{:}hasColour(X,Y)~|~colors{:}hasColour(X,Y)$ and
$colors{:}pink(X)~|~colors{:}pink(X)$. 

It seems natural to use such abstraction clauses as indexes 
to the corresponding SW~documents in a semantic search engine.
Then, the query answering process can be organised as follows.
As in the case of reasoning over RDB abstractions, a reasoner
is used to derive schematic answers to a given query, based 
on all available abstractions of indexed SW~documents. 
Each schematic answer to the query depends on some abstraction clauses.
The documents associated with these clauses are potentially
relevant to our query, so we download them, and only them, 
into our local RDB for further processing. 

Of course, the indexing scheme presented here is just a conceptual one.
The developers have the flexibility to chose a concrete representation
-- for example, they may just index by the URIs of concepts and relations, 
and only create the corresponding abstraction clauses when the reasoner 
is ready to inject them in the search space. There is also a possibility
of adjusting the degree of generality of abstraction clauses 
by adding some ad hoc constraints. For example, the first of the abstraction
clauses from the example above can be replaced with the more specific
$zoo{:}elephant(X)~|~zoo{:}elephant(X),~pref(X,''htpp://www.myelephants.com/'')$.
The ad hoc constraint $pref(X,''htpp://www.myelephants.com/'')$ requires the prefix
of the URI $X$ to be ''htpp://www.myelephants.com/''. The constraint is 
incompatible with, e.~g.,  $pref(X,''htpp://www.myrhinos.com/'')$, 
so if our reasoner derives a clause with these two constraints, 
it can safely discard it, thus improving the precision of indexing.


                 \section{Related work.}
                    \label{sec:related_work}

We are not aware of any work that uses resolution-based reasoning
in a way similar to the one proposed in this research note,
i.~e., for incremental query rewriting based on the use of 
complete query answering over database abstractions, implemented
with constraints over the concrete data.

In general, semantic access to relational databases is not 
a new concept. Some of the work on this topic is limited
to semantic access to, or semantic interpretation of relational
data in terms of Description Logic-based ontologies or RDF
(see, e.~g.,
  \cite{Calvanese+:DL:2007,Bizer:D2RQ:ISWC:2004,RanganathanLiu:SemanticQueries:ACM_CIKM:2006}
),
or non-logical semantic schemas 
(see \cite{Rishe:SemSQL}).
There is also a large number of projects and publications on 
the use of RDB for storing and querying large RDF and OWL datasets:
see, e.~g., \cite{Pan+:WPSSWS:DLDB:2003,Horrocks+:DL:InstanceStore:2004,Calvanese+:AAAI:2005,Chen+:DL:LAS:2005,Dolby+:SHINAbox:2007}, to mention just a few.
%
%
The format of the research note does not allow us
to give a comprehensive overview of such work, so we will concentrate
on research that tries to go beyond the expressivity of DL and, at the same time,
is applicable to legacy relational databases.


The work presented here was originally inspired by 
the XSTONE project \cite{Tammet+:DatabasesAndInformation:XSTONE:2006}.
In XSTONE, a resolution-based theorem prover 
(a reimplementation of Gandalf, which is, in particular, optimised
for taxonomic reasoning) is integrated with an RDBMS by loading
rows from a database as ground facts into the reasoner and using
them to answer queries with resolution. The system is highly scalable
in terms of expressiveness: it accepts full FOL with some useful
extensions, and also has parsers for RDF, RDFS and OWL.
We believe that our approach has better data scalability and 
can cope with very large databases which are beyond the reach
of XSTONE, mostly because our approach obtains answers in bulk,
and also due to the way we use highly-optimised RDBMS.

Papers \cite{OConnor+:AIME:SWTechForBiomedData:2007} and
\cite{OConnor+:RR:RDBQueryingWithOWLAndSWRL:2007} 
describe, albeit rather superficially, a set of tools 
for mapping relational databases into OWL and semantic 
querying of the RDB. Importantly, the queries are formulated as 
SWRL \cite{W3C:SWRL:URL} rule bases. Although SWRL only allows 
Horn rules built with OWL concepts, properties and equality, 
its expressivity is already sufficient for many applications.
Given a semantic query in the form of a SWRL rule base, the software generates 
SQL queries in order to extract some relevant data in the form of 
OWL assertions and runs a rule engine on this data to generate 
final answers. So the reasoning is, at least partially,
done on a per-answer basis, which gives us hope that our approach can 
scale up better.

Another project, OntoGrate \cite{Dou+:ICDE:IntegratingDBIntoSW:2006},
uses an approach to deductive query answering, which is based on 
the same ideas as ours: their FOL reasoner, 
OntoEngine \cite{Dou+:JDS:OntTransOnSemWeb:2005}, 
can be used
to rewrite original queries formulated in terms of some ontology,
into a finite set of conjunctive queries in terms of the DB schema, which 
is then converted to SQL. For this task, the reasoner uses
\emph{backward chaining with Generalised Modus Ponens} 
\cite{RusselNorvig:AIAModernApproach}, 
which corresponds
to negative hyperresolution on Horn clauses in the more common 
terminology. A somewhat ad hoc form of term rewriting 
\cite{Nieuwenhuis:HandbookAR:paramodulation:2001} is used to deal 
with equality. Termination is implemented by setting some limits
on chaining, which allows them to avoid incremental processing.
We hope to go much further, mainly, but not only,
by putting our work on a solid theoretical foundation. 
In particular, we are paying attention to completeness. 
Since our approach is based on well-studied calculi, we hope
to exploit the large amount of previous research on completeness 
and termination, which seems very difficult to do with the approach taken 
by OntoEngine. Althouth we are very likely to make various concessions
to pragmatics, we would like to do this in a controllable and reproducible
manner.

On the more theoretical side, it is necessary to mention two other connections.
The idea of using constraints to represent schematic answers 
is borrowed from Constraint Logic Programming \cite{JaffarMaher:CLP:JLP:1994} and 
Constrained Resolution \cite{BurckertNutt:QueryAnsweringWithConsRes:1992}.
Also, the general idea of using reasoning for preprocessing expressive queries 
into a database-related formalism, was borrowed from \cite{Motik:PhDThesis:2006},
where a resolution- and paramodulation-based calculus is used to translate
expressive DL ontologies into Disjunctive Datalog. This work also shares
a starting point with ours -- the observation that reasoning methods
that treat individuals/data values separately can not scale up sufficiently.

                 \section{Future work.}
                    \label{sec:future_work}


Our future work will be mostly concentrated in the following directions:

\myparagraph{Equality treatment.} If equality is present in our knowledge
bases (e.~g., in the form of OWL number restrictions), we can extend
the standard superposition calculus to clauses with recording literals
as we did with resolution. However, the completeness proof does not
easily transfer to such use of superposition. Therefore, one of our main 
priorities now is to look for adjustments of the superposition calculus 
that would be provably complete w.~r.~t. schematic answers, without
being too inefficient. An obvious obstacle to generalisation-tolerance
is the absence of paramodulations into variables in the standard paramodulation-based
calculi, so, for a start, we will try to use the specificity of reasoning 
over DB abstractions to eliminate the need for such inferences in generalisation-tolerant
variants of superposition.

\myparagraph{Completeness with redundancy deletion.} Static completeness,
proven in Section~\ref{sec:soundness_and_completeness}, is enough
to guarantee that we will find all necessary answers only if our search
procedure generates absolutely all possible derivations in the given
calculus. In practice, such approach is almost always inefficient.
Typically, some criteria are applied to detect redundant clauses and remove
them from the current clause set (see, e.~g., 
\cite{BachmairGanzinger:HandbookAR:resolution:2001,Nieuwenhuis:HandbookAR:paramodulation:2001}). 

It seems relatively easy to prove completeness of schematic answer
derivation process in presense of the most important redundancy deletion 
technique:
roughly, a clause subsumed by another clause can be deleted from the current
clause set. The main idea for such a proof is that if subsumption removes 
an answer derivation from the search space, the search space will still 
contain a structurally simpler derivation of the same answer or a more 
general answer. Note that this is a property of generalisation-tolerant
calculi. However, if we want to deal with equality efficiently, we have
to demonstrate compatibility of our approach with the \emph{standard redundancy
criterion} (see, e.~g., 
\cite{BachmairGanzinger:HandbookAR:resolution:2001,Nieuwenhuis:HandbookAR:paramodulation:2001}).

\myparagraph{Termination.} Very often it is desirable that a query answering
implementation terminates on a given query having exhausted all solutions,
e.~g., for counting and aggregation of other kinds. We are interested 
in identifying combinations 
of practically relevant fragments of FOL with reasoning methods and
strategies, that guarantee termination.

\myparagraph{Implementation and experiments.} 
A proof-of-concept implementation has been already created, 
based on a version of the \Vampire\ prover
\cite{RiazanovVoronkov:design_and_implementation_of_Vampire:AICOM:2002}, 
and two experiments 
were done -- one on a large instance of the LUBM benchmark 
\cite{Guo+:ISWC:LUBM:2004}
and another one on the BioCyc
\cite{Krummenacker:Bioinformatics:BioCyc:2005} dataset (in OWL). 
This preliminary
work will be used to guide a more substantial implementation effort 
including an implementation of a front-end for all monotonic sublanguages
of Derivation RuleML \cite{Boley+:RuleML:URL}, an implementation
of a client-server Java API and tuning the reasoner for the task
of schematic answer derivation over RDB abstractions.


\bibliographystyle{plain}
\bibliography{query_answering}

\begin{thebibliography}{10}

\bibitem{Hawke:RIFCharter:URL}
{R}ule {I}nterchange {F}ormat {W}orking {G}roup {C}harter.
\newblock \url{http://www.w3.org/2005/rules/wg/charter.html}.

\bibitem{Boley+:RuleML:URL}
{T}he {R}ule {M}arkup {I}nitiative {W}eb {S}ite.
\newblock \url{http://www.ruleml.org/}.

\bibitem{W3C:SWRL:URL}
{W3C} {SWRL} {S}ubmission:.
\newblock \url{http://www.w3.org/Submission/SWRL/}.

\bibitem{RanganathanLiu:SemanticQueries:ACM_CIKM:2006}
A.Ranganathan and Z.~Liu.
\newblock {I}nformation {R}etrieval from {R}elational {D}atabases using
  {S}emantic {Q}ueries.
\newblock In {\em Proc. ACM CIKM}, pages 820--821, 2006.

\bibitem{BachmairGanzinger:HandbookAR:resolution:2001}
L.~Bachmair and H.~Ganzinger.
\newblock Resolution {T}heorem {P}roving.
\newblock In A.~Robinson and A.~Voronkov, editors, {\em Handbook of Automated
  Reasoning}, volume~I, chapter~2, pages 19--99. Elsevier Science, 2001.

\bibitem{Bizer:D2RQ:ISWC:2004}
C.~Bizer.
\newblock {D2RQ} -- {T}reating {N}on-{RDF} {D}atabases as {V}irtual {RDF}
  {G}raphs.
\newblock In {\em Proc. of the 3rd International Semantic Web Conference (ISWC
  2004)}, 2004.

\bibitem{Boley+:RuleMLDesignRationale:SWWS:2001}
H.~Boley, S.~Tabet, and G.~Wagner.
\newblock {D}esign {R}ationale of {R}ule{ML}: {A} {M}arkup {L}anguage for
  {S}emantic {W}eb {R}ules.
\newblock In {\em Semantic Web Working Symposium (SWWS)}, 2001.

\bibitem{BurckertNutt:QueryAnsweringWithConsRes:1992}
H.-J. B{\"u}rckert and W.~Nutt.
\newblock {O}n {A}bduction and {A}nswer {G}eneration through {C}onstrained
  {R}esolution.
\newblock Technical Report DFKI RR-92-51, 1992.

\bibitem{BurhansShapiro:JAL:AnswerClasses:2007}
D.~T. Burhans and S.~C. Shapiro.
\newblock Defining {A}nswer {C}lasses {U}sing {R}esolution {R}efutation.
\newblock {\em Journal of Applied Logic}, 5:70--91, 2007.

\bibitem{Calvanese+:DL:2007}
D.~Calvanese, G.~De Giacomo, D.~Lembo, M.~Lenzerini, A.~Poggi, and R.~Rosati.
\newblock {MASTRO-I}: {E}fficient {I}ntegration of {R}elational {D}ata through
  {DL} {O}ntologies.
\newblock In {\em DL-07}, 2007.

\bibitem{Calvanese+:AAAI:2005}
D.~Calvanese, G.~De Giacomo, D.~Lembo, M.~Lenzerini, and R.~Rosati.
\newblock {DL}-{L}ite: {T}ractable {D}escription {L}ogics for {O}ntologies.
\newblock In {\em Proc. of the 20th Nat. Conf. in Artificial Intelligence (AAAI
  2005)}, pages 602--607, 2005.

\bibitem{Chen+:DL:LAS:2005}
C.~M. Chen, V.~Haarslev, and J.~Y. Wang.
\newblock {LAS}: {E}xtending {R}acer by a {L}arge {A}box {S}tore.
\newblock In {\em Proceedings of the 2005 International Workshop on Description
  Logics (DL-2005)}, pages 200--207, Edinburgh, Scotland, UK, 2005.

\bibitem{Dolby+:SHINAbox:2007}
J.~Dolby, A.~Fokoue, A.~Kalyanpur, L.~Ma, C.~Patel, E.~Schonberg, K.~Srinivas,
  and X.~Sun.
\newblock {E}fficient reasoning on large {SHIN} {A}boxes in relational
  databases.
\newblock Technical report, 2007.
\newblock (not published yet).

\bibitem{Dou+:ICDE:IntegratingDBIntoSW:2006}
D.~Dou, P.~LePendu, S.~Kim, and P.~Qi.
\newblock {I}ntegrating {D}atabases into the {S}emantic {W}eb through an
  {O}ntology-based {F}ramework.
\newblock In {\em International Workshop on Semantic Web and Database at ICDE
  2006}, pages 33--50, 2006.

\bibitem{Dou+:JDS:OntTransOnSemWeb:2005}
D.~Dou, D.~McDermott, and P.~Qi.
\newblock {O}ntology {T}ranslation on the {S}emantic {W}eb.
\newblock {\em Journal of Data Semantics}, 2:35--37, 2005.

\bibitem{Guo+:ISWC:LUBM:2004}
Y.~Guo, J.~Heflin, and Z.~Pan.
\newblock An {E}valuation of {K}nowledge {B}ase {S}ystems for {L}arge {OWL}
  {D}atasets.
\newblock In {\em Third International Semantic Web Conference (ISWC 2004)},
  pages 613--627, 2004.

\bibitem{Horrocks+:DL:InstanceStore:2004}
I.~Horrocks, L.~Li, D.~Turi, and S.~Bechhofer.
\newblock {T}he {I}nstance {S}tore: {D}escription {L}ogic {R}easoning with
  {L}arge {N}umbers of {I}ndividuals.
\newblock In {\em International Workshop on Description Logics (DL 2004)},
  2004.

\bibitem{JaffarMaher:CLP:JLP:1994}
J.~Jaffar and M.~J. Maher.
\newblock {C}onstraint {L}ogic {P}rogramming: a {S}urvey.
\newblock {\em Journal of Logic Programming}, 19(20):503--581, 1994.

\bibitem{Krummenacker:Bioinformatics:BioCyc:2005}
Markus Krummenacker, Suzanne Paley, Lukas Mueller, Thomas Yan, and Peter~D.
  Karp.
\newblock {Q}uerying and {C}omputing with {B}io{C}yc {D}atabases.
\newblock {\em Bioinformatics}, 21(16):3454--3455, 2005.

\bibitem{Motik:PhDThesis:2006}
B.~Motik.
\newblock {\em Reasoning in {D}escription {L}ogics using {R}esolution and
  {D}eductive {D}atabases}.
\newblock Ph{D} {T}hesis, Karlsruhe University, Karlsruhe, January 2006.

\bibitem{Nieuwenhuis:HandbookAR:paramodulation:2001}
R.~Nieuwenhuis and A.~Rubio.
\newblock Paramodulation-{B}ased {T}heorem {P}roving.
\newblock In A.~Robinson and A.~Voronkov, editors, {\em Handbook of Automated
  Reasoning}, volume~I, chapter~7. Elsevier Science, 2001.

\bibitem{OConnor+:RR:RDBQueryingWithOWLAndSWRL:2007}
M.J. O'Connor, R.D. Shankar, S.W. Tu, C.~Nyulas, A.K. Das, and M.A. Musen.
\newblock {E}fficiently {Q}uerying {R}elational {D}atabases {U}sing {OWL} and
  {SWRL}.
\newblock In {\em Web Reasoning and Rule Systems,First International
  Conference, RR 2007}, pages 361--363, 2007.

\bibitem{OConnor+:AIME:SWTechForBiomedData:2007}
M.J. O'Connor, R.D. Shankar, S.W. Tu, C.~Nyulas, D.B. Parrish, M.A. Musen, and
  A.K. Das.
\newblock {U}sing {S}emantic {W}eb {T}echnologies for {K}nowledge-{D}riven
  {Q}uerying of {B}iomedical {D}ata.
\newblock In {\em 11th Conference on Artificial Intelligence in Medicine (AIME
  07)}, 2007.

\bibitem{Pan+:WPSSWS:DLDB:2003}
Z.~Pan and J.~Heflin.
\newblock {DLDB}: {E}xtending {R}elational {D}atabases to {S}upport {S}emantic
  {W}eb {Q}ueries.
\newblock In {\em Workshop on Practical and Scaleable Semantic Web Systems,
  ISWC 2003}, pages 109--113, 2003.

\bibitem{Ramakrishnan:Database_Management_Systems:2003}
R.~Ramakrishnan and J.~Gehrke.
\newblock {\em {D}atabase {M}anagement {S}ystems, third edition}.
\newblock McGraw-Hill, 2003.

\bibitem{RiazanovVoronkov:design_and_implementation_of_Vampire:AICOM:2002}
A.~Riazanov and A.~Voronkov.
\newblock The {D}esign and {I}mplementation of {Vampire}.
\newblock {\em AI Communications}, 15(2-3):91--110, 2002.

\bibitem{Rishe:SemSQL}
N.~Rishe.
\newblock {S}emantic{SQL}: {A} {S}emantic {W}rapper for {R}elational
  {D}atabases.
\newblock \url{http://n1.cs.fiu.edu/semantic.wrapper.pdf}, 2004.
\newblock (white paper).

\bibitem{RusselNorvig:AIAModernApproach}
S.~Russel and P.~Norvig.
\newblock {\em {A}rtificial {I}ntelligence: {A} {M}odern {A}pproach, {S}econd
  {E}dition}.
\newblock Prentice-Hall, Inc., 2003.

\bibitem{Tammet:JAR:Gandalf:1997}
T.~Tammet.
\newblock Gandalf.
\newblock {\em Journal of Automated Reasoning}, 18(2):199--204, 1997.

\bibitem{Tammet+:DatabasesAndInformation:XSTONE:2006}
T.~Tammet, V.~Kadarpik, H.-M. Haav, and M.~Kaaramees.
\newblock A {R}ule-based {A}pproach to {W}eb-based ({D}atabase) {A}pplication
  {D}evelopment.
\newblock In {\em 7th International Baltic Conference on Databases and
  Information Systems}, pages 202--208, 2006.

\end{thebibliography}

\end{document}